# Multilevel active registration for kinect human body scans: from low quality to high quality

Zongyi Xu[1] · Qianni Zhang[1] · Shiyang Cheng[2]

**Abstract** Registration of 3D human body has been a challenging research topic for over decades. Most of the traditional human body registration methods require manual assistance, or other auxiliary information such as texture and markers. The majority of these methods are tailored for high-quality scans from expensive scanners. Following the introduction of the low-quality scans from cost-effective devices such as Kinect, the 3D data capturing of human body becomes more convenient and easier. However, due to the inevitable holes, noises and outliers in the low-quality scan, the registration of human body becomes even more challenging. To address this problem, we propose a fully automatic active registration method which deforms a high-resolution template mesh to match the low-quality human body scans. Our registration method operates on two levels of statistical shape models: (1) the first level is a holistic body shape model that defines the basic figure of human; (2) the second level includes a set of shape models for every body part, aiming at capturing more body details. Our fitting procedure follows a coarse-to-fine approach that is robust and efficient. Experiments show that our method is comparable with the state-of-the-art methods for high-quality meshes in terms of accuracy and it outperforms them in the case of low-quality scans where noises, holes and obscure parts are prevalent.

## 1 Introduction

The modeling of accurate 3D human body is a fundamental problem for many applications such as design, animation, and virtual reality. The modeling of human body meshes is performed on a corpus of registered scans. However, the acquirement of high-quality human body meshes and registration of meshes are challenging. Current publicly available high-quality human body datasets, such as SCAPE [3], FAUST [5], TOSCA [8] are built either from costly laser scanners or need other assistance (e.g makers, texture or professional tools). With the appearance of low-cost scanners such as Kinect, it is now possible for an object, a room or even a person to be quickly scanned, modeled and tracked [12, 14, 27, 28, 30, 31, 44]. Nowadays human body meshes could be captured for different identities in different poses in a few minutes. However, the prevalent noises, outliers and holes in the scans acquired with low-cost scanners bring in more challenges for mesh registration.

To register the 3D scans, several 3D fitting methods are proposed [1, 2, 5, 14, 32, 49]. The invertible finite volume method [14] is used to control the template tetrahedral mesh to the target point clouds. The stitched puppet model [49] adopts the DPMP algorithm which is a particle-based method to align a graphical model to target meshes. More efforts are made to perform the nonrigid ICP (iterative closest point) [1, 2] which computes the

✉ Zongyi Xu
zongyi.xu@qmul.ac.uk

Qianni Zhang
qianni.zhang@qmul.ac.uk

Shiyang Cheng
shiyang.cheng11@imperial.ac.uk

[1] Queen Mary University of London, Mile End Rd, London E1 4NS, UK

[2] Imperial College London, Kensington, London SW7 2AZ, UK

affine transformation at each vertex of template to allow non-rigid registration of template and scans. Although these ICP-based nonrigid registration methods demonstrate high accuracy, it is sensitive to missing data, which might lead to an erroneous fitting result. For Kinect-like scanners, due to self-occluded parts like crotch and armpit, holes and distortion on the mesh are inevitable.

To faithfully register the body scans captured from low-cost scanners, like Kinect, we present a multilevel active body registration (MABR) approach to build a watertight and high fidelity virtual human body in an automatic way. We aim to align a template mesh with the target scans acquired with Kinect as close as possible. Here, a template mesh is the mean shape which is learned from an existing high-quality human body mesh dataset. In our method, multilevel registration is performed. In the first level, the overall template and target are roughly aligned. In the second level, a region-based registration is performed where the template is divided into 16 parts and each part is fitted to the target separately. For the main body parts where the scan is complete and full of details such as torso, legs and arms, the local affine transformation for each vertex is computed. As for impaired parts such as foot and hand, we deform the corresponding parts of the template at a coarse-grained level for completeness.

With the proposed method, we are able to automatically reconstruct high-quality 3D mesh from low-quality scans or point clouds. This technique can be employed in a variety of applications such as in virtual shopping applications to show the clothes from different stereo views and help the customers to choose the best fitting clothes. In the virtual games, the systems can generate realistic full body avatars according to rough scans of the users instantly, which benefit from algorithm's robustness to missing data which commonly exist in scans from low-cost scanners. The approach manages to avoid the tediously manual work of building high-fidelity 3D models with professional tools and is capable of building a complete and high-quality meshes within 2 min automatically, which can be beneficial to the television production. This method may also be integrated in software as a tool for preprocessing raw scans, filling in missing parts automatically and registering scans.

Our main contributions reported in this paper are:

- First, we propose a fully automatic registration method which performs well even on noisy low-quality data. Our method follows the region-based approach to register the human body scans, which improves the accuracy of registration. According to the nature of different body parts, our approach adopts particular registration strategies, which makes the method robust to noisy Kinect scans.

- Second, we provide a dataset of 250 real human body scans acquired with Microsoft Kinect for XBOX 360. This dataset can be used to evaluate the robustness of registration algorithms in case of low-quality scans. The dataset is available for research purposes at http://www.eecs.qmul.ac.uk/~zx300/k3d-hub.html.

The rest of this paper is structured as follows. In Sect. 2, the literature review of mesh registration is presented. Our proposed method is described in detail in Sect. 3 and we also introduce the Kinect scanning platform which is used to build our K3D-Hub dataset in Sect. 4. The experimental evaluation results are shown in Sect. 5 and a brief summary is given in Sect. 6.

## 2 Related work

Although shape matching has been deeply researched, finding full correspondences for non-rigid and articulated meshes is still challenging. Geometry information is usually used to extract local features. Histogram of Oriented Normal Vectors [40] and Local Normal Binary Patterns (LNBPs) [37] are descriptors presented based on surface normal. Since the colour information cannot represent the unique feature in 3D mesh domain, it usually is used as an auxiliary information to other features [5]. Besides using the local geometric features, many works extend the existing 2D features to the 3D domain [13, 33, 38]. 3D-Harris [33] is the 3D extension of the 2D corner detection method with Harris operator. Local depth SIFT (LD-SIFT) [13] extends SIFT feature by representing the vicinity of each interest point as a depth map and estimating its dominant angle using the principal component analysis to achieve rotation invariance. MeshSIFT [38] characterizes the salient points neighbourhood with a feature vector consisting of concatenated histograms of shape indices and slant angles. MeshSIFT presents robustness to expression variations, missing data and outliers when it is used to 3D face shape matching. Clearly, both of these methods rely on the local shape features such as curvature or angles. Since they are not pose independent, they cannot be used for shapes undergoing affine transformation, like human body shape with different poses.

Since human body is isometric shape, many works make use of isometry to find the correspondences. If two shapes are perfectly isometric, then there exists an isometry i.e., a distance-preserving mapping, between these shapes such that the geodesic distance between any two points on one shape is exactly the same as the geodesic distance between their correspondences on the other [36]. Different approaches are proposed to exploit isometry for shape correspondences [14, 15, 20, 29, 35]. One way is to embed shape into a different domain where geodesic distances are replaced by Euclidean distance so that isometric deviation can be measured and optimized in the embedding space [15]. Euclidean embedding can be achieved using various techniques such as classical MDS (Multidimensional Scaling) [20, 35], least-squares MDS [15], and spectral analysis of the graph Laplacian [29] or of the Laplace–Beltrami operator [14]. However, when it comes to the meshes from low-cost scanners, the above isometry-based methods are not applicable as they usually require watertight meshes and suffer from self-symmetry of human body shape.

Another approach is to fit a common template mesh to noisy scans. Once fitted, these scans share a common topology with the template and are fully registered. By removing noises and completing holes in the low-quality scans, a high-quality mesh is built straightforwardly. To perform registration, traditional methods tend to rely on auxiliary modeling tools, such as Maya,[1] Blender,[2] manual markers and texture information. Recently, authors in [26] deform a high-quality template mesh to scans which are from a stereo scanning system consisting of multiple RGB-D cameras in a circle. Various non-rigid ICP algorithms [2, 16, 17, 21, 22, 24] are proposed to register 3D mesh. They usually combine the classic ICP with some regularization terms to make the surface deformation smooth. However, the ICP-based methods are sensitive to missing data and outliers. When they are used in noisy Kinect scans, the hand/foot parts and top of the head are usually distorted severely.

Besides the ICP-based registration methods mentioned above, statistical shape models are employed to improve the smoothness and robustness, as the prior knowledge are embedded. Scape [3] learns a shape model with PCA to describe the body shape variations using 45 instances in a similar pose. It also builds a pose model which is a mapping from posture parameters to the body shape with a dataset that includes 70 poses of one subject. With the learnt model, it builds a human body dataset but only pose dataset is released which contains meshes of 70 different poses of a particular person. Since the body shapes of different people vary greatly for a particular pose (for example, considering the same pose of arm lifting, the muscle variations of normal people and the athlete are definitely different.), TenBo [9] proposes to model 3D human body with variations on both pose and body shape. It trained the Tenbo model with the dataset from [18]. The model is used to estimate shape and pose parameters with the depth map and skeleton provided by Microsoft Kinect sensors. The FAUST [5] contains 300 scans of 10 people in 30 different poses. The authors make use of the texture information to assist the alignment of the meshes. The registered mesh has 6890 vertices and 13,776 faces. Compared with SCAPE dataset, the resolution is lower but the mesh is still realistic. Nonetheless, its registration method is not fully automatic for the reason that it is based on the texture information which is added by hand. The CAESAR dataset [34] contains 2400 male and female laser scans with texture information and hand-placed landmarks. Each range scan in the dataset has about 150,000–200,000 vertices and 73 markers. Unfortunately, this dataset does not provide correspondences and contains many holes. The MPI [18] captures 114 subjects in a subset of 35 poses using a 3D laser scanner. All the aforementioned models are captured from expensive scanners or under the condition of complex and large scale scanning platform. Compared with scans acquired with low-cost scanners, they have much less noises floating on the surface, no big holes and no hierarchical outliers. The methods working on these high-quality meshes might not be directly applied to low-quality scans from cheap scanners, like Kinect, to get satisfactory results.

The statistical shape model also has been introduced for 3D face reconstruction, face modeling and face animation [4, 41–43]. Unlike human body, the facial landmarks [21, 45–47] can be detected accurately and used as reliable constraints to initialize the fitting of morphable model. In [25], for aligning two faces, the authors extract the facial features before performing ICP registration. Accurate landmarks are extracted in [7] to guide the face modeling from large-scale facial dataset. In [23], a pre-processing algorithm is proposed to fill holes and smooth the noisy depth data from Kinect before performing face recognition. A high-resolution face model is constructed in [6] from low-resolution depth frames acquired with a Kinect sensor. In this work, an initial denoising operation which is based on the anisotropic nature of the error distribution with respect to the viewing direction of the acquired frames and a following manifold estimation approach based on the lowess nonparametric regression method which is used to remove outliers from the data are proposed to generate high-resolution face models from Kinect depth sequences. However, these approaches detect landmarks with the help of RGB images or depth images as clear and strong initialization or preprocessing steps are performed to fill holes or smooths the data. In the case of human body registration where texture information is often missing, accurate initial landmarks are hard to be detected automatically. Compared with human faces, the magnitude of changes of human body surface is larger even though the subjects are asked to perform the same pose, which brings in more challenges in human body registration.

---

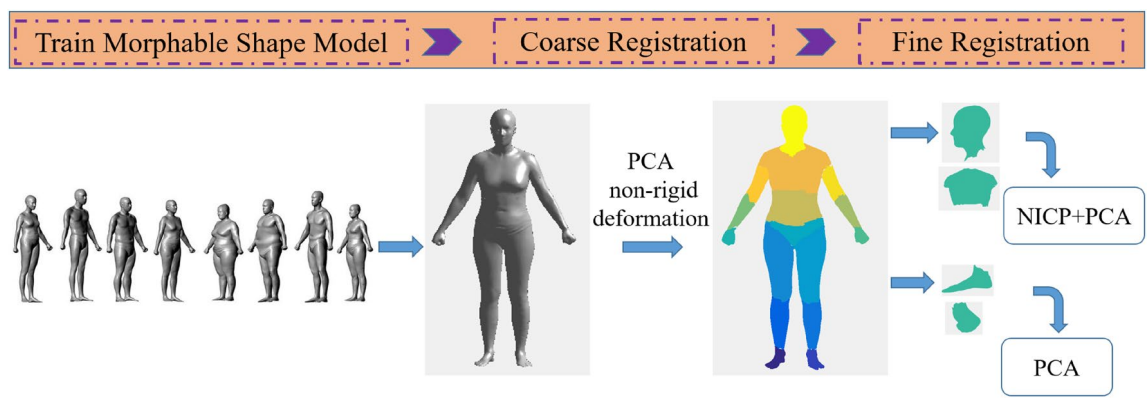

**Fig. 1** The work flow of the proposed method. We first train a statistical shape model from 200 aligned meshes in SPRING dataset using PCA techniques. The mean shape is used as the template mesh. The registration between template and target mesh includes two levels illustrated as follows. In the coarse registration level, we deform the template mesh non-rigidly into the target, making the template overlap with target in most parts. In the fine registration level, a region-based deformation is used to deform the template more accurately

## 3 Region-based human body registration

Region-based modeling technique has been prevalent in face [10, 41] and human body [49] modelling, as it allows for richer shape representation and enables the fitting of different parts to be specifically tailored. Inspired by [10], we combine the statistical shape model with non-rigid iterative closest point algorithm. However, the direct application of this fitting method to low-cost, noisy and incomplete Kinect scans could lead to inconsistent and erroneous results. This happens particularly often when it comes to hands and feet fitting (examples of failed fitting shown in Fig. 9). The main reason is that Kinect scan of the feet can barely be separated from the stand; while, during data capturing, negligible movement of hands is inevitable, causing serious artifacts in hand scan. Even if we perform the coarse level registration, the distance of these parts between source and target might be large, the nearest neighbors tend to be incorrect and non-rigid ICP easily gets trapped in local minima [19]. Therefore, we propose a different fitting method that takes special care of foot and hand modeling. The pipeline of the proposed MABR method is shown in Fig. 1. First, a 3D morphable shape model is trained from 200 pre-aligned high-quality meshes. The mean shape is used as template. Second, coarse registration is employed to roughly align the template and target. Then different non-rigid deformation techniques are applied on the main body parts and hand/foot parts, respectively, with our trained morphable shape model.

### 3.1 Rigid registration

The target mesh is captured from a Kinect scanner and the template mesh is the mean shape from the public dataset. The goal of rigid registration is to unify their coordinate systems. In traditional rigid transformation, correspondences are needed to compute the rigid transformation matrix. Some works use markers to establish correspondences manually. Some 3D mesh features like Heat Kernel Signature [39] are based on surface properties like geodesic distance, curvature, or face normals. These features work well on public human mesh dataset as they are processed to share topology and high-quality without noises or folding faces so that the geometry distance is measurable. However, in our case, the number of vertices of targets varies while the template mesh has fixed number of vertices. Moreover, it is obvious that the physique such as height and muscle properties of the template are different from those of scans. Lastly, noises and holes exist in our data. Therefore, the feature which works on the high-quality surface cannot be used in our work.

Without using correspondence, we choose to build a shape-aware coordinate system for each model and transform the source to align its origin and axes with the target. PCA is used to identify the most important parts from the vertex set. PCA-based alignment is to align the principle directions of the vertex set. First, given a set of vertices $S_p = \{p_i\}$ and its centroid location $\mathbf{c}$, we have $\mathbf{P}$ formulated as Eq. 1.

$$\mathbf{P} = \begin{bmatrix} p_{1x} - c_x, p_{2x} - c_x, \ldots, p_{nx} - c_x \\ p_{1y} - c_y, p_{2y} - c_y, \ldots, p_{ny} - c_y \\ p_{1z} - c_z, p_{2z} - c_z, \ldots, p_{nz} - c_z \end{bmatrix}, \quad (1)$$

where $p_{ix}, p_{iy}, p_{iz}$ and $c_x, c_y, c_z$ are the coordinates of vertex $\mathbf{p_i}$ and centroid $\mathbf{c}$ respectively. The covariance matrix $\mathbf{M}$ is formulated as:

$$\mathbf{M} = \mathbf{PP}^T \quad (2)$$

The eigenvectors of the covariance matrix $\mathbf{M}$ represent principle directions of shape variation. They are orthogonal to each other while the eigenvalues indicate the amount of variation along each eigenvector. Therefore, the eigenvector with largest eigenvalue is the direction where the mesh shape varies the most. In the human body mesh, the principle direction should be along the height direction. The next two directions should be along the width and thickness of the human body, respectively.

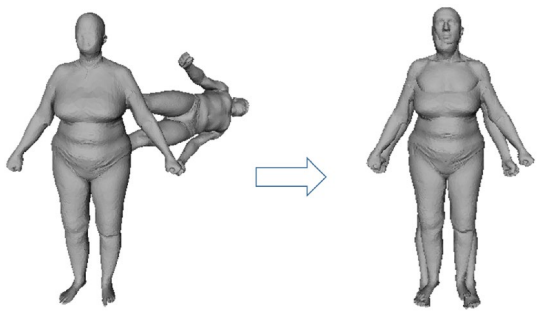

Given two human body meshes $S_a$ and $S_b$, their covariance matrices $\mathbf{M_a}$ and $\mathbf{M_b}$ can be computed with Eq. 2. We form two matrices $\mathbf{E_a}$ and $\mathbf{E_b}$ where columns are the eigenvectors of $\mathbf{M_a}$ and $\mathbf{M_b}$, respectively. To align these two orthogonal matrices, we compute the rotation $\mathbf{R}$ such that

$$\mathbf{RE_a} = \mathbf{E_b}, \tag{3}$$

Finally, the PCA-based alignment can be performed with the following formula.

$$\mathbf{S_b}' = \mathbf{c_a} + \mathbf{R}(\mathbf{S_b} - \mathbf{c_b}), \tag{4}$$

where $\mathbf{c_a}$ and $\mathbf{c_b}$ are the centroids of $S_a$ and $S_b$ correspondingly. After we perform the rigid registration, $S_b'$ should be aligned with $S_a$ in terms of main directions. One of the initially rigid alignment examples is shown in Fig. 2. We can see that both of meshes look forward after we align them rigidly. However, in terms of height, body shapes, they still differ a lot.

### 3.2 Morphable shape models

In this part, we introduce the statistical body shape model trained from 200 entire human body meshes using PCA technique. The training set is from the SPRING dataset [48] which includes 3038 high-resolution body models and each mesh has 12,500 vertices and 25,000 faces. All the meshes have been placed in point to point correspondence. This large aligned dataset allows for a reliable model to be learnt robustly. Given a set of training shapes, the statistical shape model can be represented as:

$$\mathbf{v} = \mathbf{Bc} + \mathbf{m}, \tag{5}$$

where $\mathbf{v} \in \mathfrak{R}^{4N \times 1}$ are the 3D coordinates $(x, y, z)$ plus corresponding homogeneous coordinates of all $N$ vertices; $\mathbf{B} \in \mathfrak{R}^{4N \times k}$ are the eigenvectors of the PCA model, $\mathbf{m} \in \mathfrak{R}^{4N \times 1}$ is the mean shape, and $\mathbf{c} \in \mathfrak{R}^{k \times 1}$ contains the non-rigid parameters for shape deformation.

Apart from a holistic body shape model, to further describe the large amount of shape variability in human body, we model each region of the body with its own PCA model. In this paper, we employ the body segmentation model provided by the SCAPE [3] dataset. Assume that we have $p$ independent parts in the segmented template $\mathcal{V} = \{\mathbf{v}^i\}_{i=1}^p$, and the $i$th part $\mathbf{v}_i$ can also be modeled using Eq. 6:

$$\mathbf{v}^i = \mathbf{B}^i \mathbf{c}^i + \mathbf{m}^i. \tag{6}$$

Here, $\mathbf{v}^i$, $\mathbf{B}^i$ and $\mathbf{m}^i$ are the shape coordinates, eigenbasis and mean shape of the model for $i$th region, respectively, and $\mathbf{c}^i$ is the latent variable controlling deformation of the model. As a result, we trained two levels of shape model: the first level is a holistic model for the entire body and the second ones is region-based model that models each body part separately.

### 3.3 Coarse level registration

The main goal of this registration is to overlap the template and target scan, while minor details of the body can be ignored in this level. After rigid transformation, we apply the holistic PCA model trained in the Sect. 3.2 to get the deformed template that would sit closer to the target point clouds. Here, with target point clouds $\mathbf{u}$ retrieved by nearest neighbors search using the k-d tree algorithm, the cost function to be minimized can be formulated as:

$$E(\mathbf{c}) = ||\mathbf{v} - \mathbf{u}||^2 = ||(\mathbf{Bc} + \mathbf{m}) - \mathbf{u}||^2. \tag{7}$$

To solve this equation, we take the partial derivative with regard to $\mathbf{c}$ and take the minimum when it approaches to zero:

$$\mathbf{B}^T \mathbf{Bc} + \mathbf{B}^T(\mathbf{m} - \mathbf{u}) = \mathbf{0}, \tag{8}$$

and get the closed-form solution,

$$\mathbf{c} = -(\mathbf{B}^T \mathbf{B})^{-1} \mathbf{B}^T(\mathbf{m} - \mathbf{u}). \tag{9}$$

### 3.4 Fine level registration

After the coarse level registration, to capture the non-rigid nature of body surface and provide an accurate fitted mesh, we make use of the region-based statistical shape model described in Sect. 3.2 and combine it with non-rigid iterative closest points (NICP) algorithm [2]. Note that during scanning, the subject is unlikely to hold the exact pose like template, especially in the parts of arm and leg, thus the hand and foot parts could easily appear as outliers.

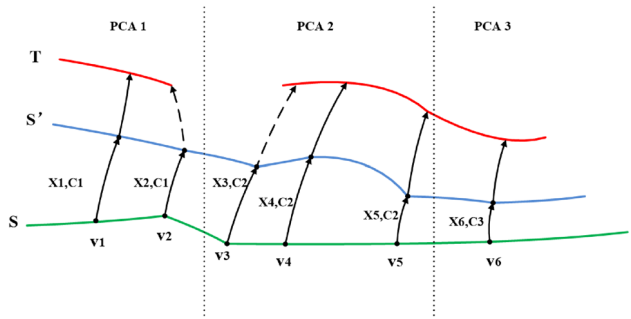

**Fig. 3** The summary of our matching framework. Our target is to find a set of affine transformations $X_i$ and local PCA parameters $C_i$, such that, when applied to the vertices $v_i$ of the template mesh $S$, result in a new surface $S'$ that matches the target surface $T$. This diagram shows the match in progress; $S'$ is moving towards the target but has not reached it. The whole vertices are divided into three parts which are controlled by three local PCAs. The transformation of each vertex is controlled by affine transformation as well as the local parameters of the part which the vertex belongs to

Although the first level fitting alleviates this effect, original NICP algorithm still might not generate satisfactory fitting result. Therefore, for the parts of hand and foot, we use only the non-rigid parameters to control the deformation in a coarse grained level and add a regularization term to make it smooth on the boundary. In this way, we can recover the hand and foot parts which are impaired in the scanning process. The clear and semantic hands and feet allow for the shape statistical modeling in the next stage.

### 3.4.1 Main body registration

We define the body parts that exclude feet and hands as the *main body*. For the *main body* parts, we combine the statistical shape model with NICP algorithm. Our goal is to find a set of affine matrices $X = \{X^i\}_{i=1}^p$ and non-rigid parameters $C = \{c^i\}_{i=1}^p$ such that the sum of Euclidean distances between pair of points of each region is minimal. Here, $X^i$ is a $3 \times 4n_i$ matrix that consists of affine matrix for every template vertex in the $i$th part. As shown in Fig. 3, we describe our technique for fitting a template $S$ to target mesh $T$. Each of these surface is represented as a triangle mesh. Each vertex $v_i$ is influenced by a $4 \times 3$ affine matrix $X_i$ and non-rigid parameter $C_i$. We define data error with these two parameters. The data error, indicated by the arrows in Fig. 3, is a weighted sum of the squared distances between template surface $S'$ and target surface $T$. Besides data error, to deform the template smoothly, we also define a stiffness term to constraint the vertices, which do not move directly towards the target, but may move parallelly along it. These error terms are summarized in Fig. 3 and described in detail in the following.

*Distance term* The distance term is used to minimize the Euclidean distances between source and target. We assume each part has $n_i$ points and the cost function is denoted as the sum of error of each pair of vertices:

$$E_d(\mathbf{X}) = \sum_{i=1}^p \sum_{j=1}^{n_i} ||\mathbf{X}_j^i \mathbf{v}_j^i - \mathbf{u}_j^i||^2, \tag{10}$$

where $X_j^i$ is the transformation matrix for $j$th vertex in the $i$th part.

Since each part is modeled by the shape model $\mathbf{v}_j^i = \mathbf{B}_j^i \mathbf{c}_j^i + \mathbf{m}_j^i$, based on Eq. 6, the distance term could be rewritten and rearranged as:

$$
\begin{aligned}
E_d(\mathbf{X}) &= \sum_{i=1}^p \sum_{j=1}^{n_i} ||\mathbf{X}_j^i(\mathbf{B}_j^i \mathbf{c}_j^i + \mathbf{m}_j^i) - \mathbf{u}_j^i||_F^2 \\
&= \sum_{i=1}^p \left\| \begin{bmatrix} \mathbf{X}_1^i & & \\ & \ddots & \\ & & \mathbf{X}_{ni}^i \end{bmatrix} \begin{bmatrix} \hat{\mathbf{v}_1^i} \\ \vdots \\ \hat{\mathbf{v}_{ni}^i} \end{bmatrix} - \begin{bmatrix} \mathbf{u}_1^i \\ \vdots \\ \mathbf{u}_{ni}^i \end{bmatrix} \right\|_F^2.
\end{aligned}
\tag{11}
$$

We can see that the above equation is not in the standard linear form of $\mathbf{AX} - \mathbf{B} = \mathbf{C}$. To differentiate, we need to swap the position of the unknown $\mathbf{X}$ and $\mathbf{V} = [\hat{\mathbf{v}_1^i}, \dots, \hat{\mathbf{v}_{n_i}^i}]^T$. Therefore, we obtain the following form.

$$E_d(\mathbf{X}) = \sum_{i=1}^p ||\mathbf{D}^i \mathbf{X}^i - \mathbf{U}^i||^2, \tag{12}$$

where the term $\mathbf{D}^i = diag({\mathbf{v}_1^i}^T, {\mathbf{v}_2^i}^T, ..., {\mathbf{v}_{n_i}^i}^T)$, and the set of closest points $\mathbf{U}^i = [\mathbf{u}_1^i, \mathbf{u}_2^i, ..., \mathbf{u}_{n_i}^i]^T$.

*Stiffness term* The stiffness term penalizes the difference between the transformation matrices of neighboring vertices. Similar to [2], it is defined as:

$$E_s(\mathbf{X}) = \sum_{i=1}^p ||(\mathbf{M}^i \otimes \mathbf{G}^i)\mathbf{X}^i||_F^2, \tag{13}$$

here, for the $i$th body part, $\mathbf{G}^i = (1, 1, 1, \gamma^i)$, where $\gamma^i$ is used to balance the scale of rotational and skew factor against the translational factor. It depends on the units of the data and the deformation type to be expressed. $\mathbf{M}^i$ is the node-arc incidence matrix of the template mesh topology [2].

*Complete cost function:* We combine Eqs. 11 and 13 to obtain the complete cost function:

$$
\begin{aligned}
E(\mathbf{X}) &= E_d(\mathbf{X}) + E_s(\mathbf{X}) \\
&= \sum_{i=1}^p \left\| \begin{bmatrix} \mathbf{D}^i \\ \mathbf{M}^i \otimes \mathbf{G}^i \end{bmatrix} \mathbf{X}^i - \begin{bmatrix} \mathbf{0} \\ \mathbf{U}^i \end{bmatrix} \right\|_F^2.
\end{aligned}
\tag{14}
$$

Equation 14 is not a quadratic function and it is difficult to obtain the optimal local affine transformation $\mathbf{X}^i$ and non-rigid parameters $\mathbf{c}^i$ simultaneously. In [10], an alternating optimization scheme is employed to solve this problem. In this paper, we use the same optimization method to find the optimal set of parameters. For details of the solution, please refer to [11].

### 3.4.2 Hands and feet registration

Although the main body parts are roughly aligned after the first level registration, the distance between source hands/feet and corresponding target is large in most cases. In this situation, the ICP-based methods easily get trapped in local minima [19]. To address this problem, we perform a PCA-based fitting for the individual part of hand/foot. Given one particular part model of hand/foot that has eigenbasis $\mathbf{B}^*$ and mean shape $\mathbf{m}^*$, we define our objective function that consists of a distance term and a regularization term, and try to obtain the optimal non-rigid parameters $\mathbf{c}^*$ by minimizing it.

*Distance term* It is defined similar to Eq. 11, but without the affine transformation matrix,

$$E_d(\mathbf{c}^*) = ||(\mathbf{B}^*\mathbf{c}^* + \mathbf{m}^*) - \mathbf{u}^*||_F^2. \tag{15}$$

*Boundary smoothness term* To stitch hand/foot with its neighboring part smoothly, we define a boundary smoothness term as follows:

$$E_b(\mathbf{c}^*) = ||\mathbf{S}^*(\mathbf{B}^*\mathbf{c}^* + \mathbf{m}^*) - \mathbf{F}^*||_F^2, \tag{16}$$

where $\mathbf{S}^*$ is the selection matrix of hand/foot parts that picks out the boundary points. $\mathbf{F}^*$ is the boundary points of the neighboring part. By enforcing the boundary constraints between two parts, we can regulate the part fitting process to avoid erroneous result caused by outlier.

*Complete cost function* The fitting objective function can be formulated as:

$$
\begin{aligned}
E(\mathbf{c}^*) &= \alpha E_d(\mathbf{c}^*) + (1 - \alpha)E_b(\mathbf{c}^*) \\
&= \left\| \begin{bmatrix} \alpha\mathbf{I} \\ (1 - \alpha)\mathbf{S}^* \end{bmatrix} (\mathbf{B}^*\mathbf{c}^* + \mathbf{m}^*) - \begin{bmatrix} \alpha\mathbf{u}^* \\ (1 - \alpha)\mathbf{F}^* \end{bmatrix} \right\|_F^2 \\
&= \left\| \mathbf{A}^*(\mathbf{B}^*\mathbf{c}^* + \mathbf{m}^*) - \mathbf{U}^* \right\|_F^2,
\end{aligned}
\tag{17}
$$

where $\alpha$ is the weighting factor between two terms, $\mathbf{A}^* = [\alpha\mathbf{I}, (1 - \alpha)\mathbf{S}^*]^T$ and $\mathbf{U}^* = [\alpha\mathbf{u}^*, (1 - \alpha)\mathbf{F}^*]^T$. This is a well-known linear least square problem. The minimum occurs where the gradient vanishes, that is $\partial E_{\mathbf{c}^*}/\partial \mathbf{c}^* = \mathbf{0}$. Thus, Eq. 17 has closed-form solution:

$$\mathbf{c}^* = -[(\mathbf{A}^*\mathbf{B}^*)^T(\mathbf{A}^*\mathbf{B}^*)]^{-1}(\mathbf{A}^*\mathbf{B}^*)^T(\mathbf{A}^*\mathbf{m}^* - \mathbf{U}^*). \tag{18}$$

The proposed fitting method for hand and feet has a nice convergence property, we show one example of residual error curve for all iterations of fitting in Fig. 11.

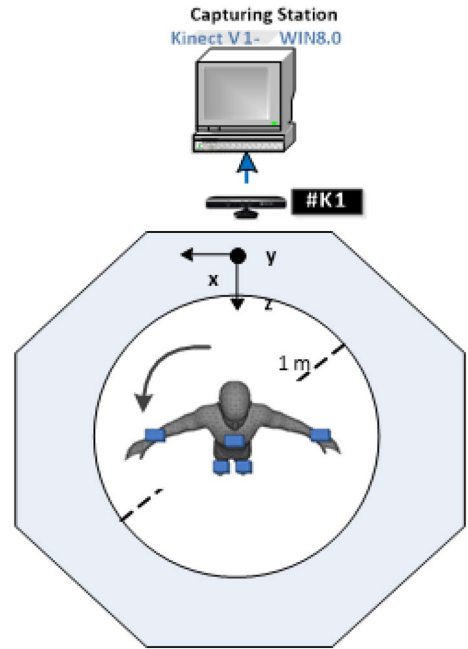

**Fig. 4** The top view of spatial arrangement of offline 3D capturing platform

## 4 Kinect scanning platform

In this part, we introduce the 3D scanning platform with single Microsoft Kinect for Xbox 360. The setup is shown in Fig. 4. The platform is built upon ReconstructMe application[3] which is based on Kinect Fusion [30].

To get the best mesh, we choose to keep the Kinect position still at three different heights when the subject is standing on a running turntable at a certain speed (30 s per round). After we scan one round at the first height, we adjust the height of the Kinect to the second height and scan the second round around the subject. The self-occluded parts such as armpit and crotch are rescanned if the Kinect does not see them in the first time. For each mesh, from our experience, it takes about 90 s to build with this platform. During data capturing, we require the participants to wear tight clothes. Each person is captured 5 poses which include a natural pose, and other 4 poses (The pose examples are shown in Fig. 13). The capturing process is displayed in Fig. 5.

Although some occluded parts can be rescanned, holes still exist on the top of the head and the soles of the feet which the Kinect cannot see. To the best of our knowledge, there is no public Kinect-based human body mesh dataset. Therefore, we utilise the platform above to build a low-quality mesh dataset, named Kinect-based 3D Human Body (K3D-Hub) Dataset. So far, our K3D-Hub dataset contains 50 different identities and 5 poses for each person. We show examples of our dataset in Fig. 13.

---

[3] http://reconstructme.net/.

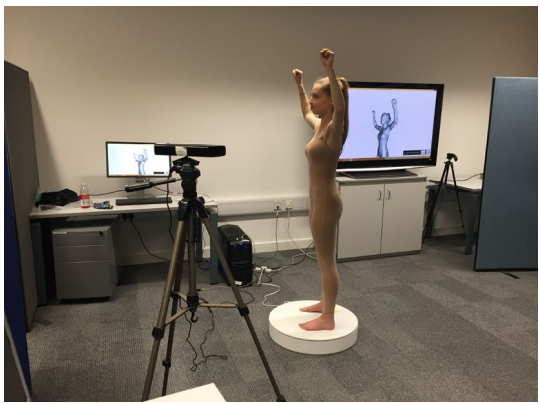

**Fig. 5** The screenshot of our capturing process

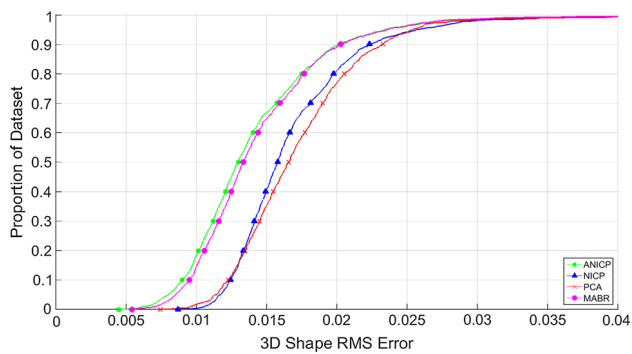

**Fig. 6** The comparison of 3D shape RMS error of ANICP, NICP, PCA and our MABR

## 5 Performance evaluation

To evaluate the performance of our MABR method, we conducted experiments on both high- and low-quality meshes, and showed the shape root mean square (RMS) error curve as well as some fitting results for visualization purpose.

### 5.1 High-quality mesh evaluation

For the evaluation on high-quality data, we use the SPRING [48] dataset that contains 3038 meshes with various human body shapes. These good quality meshes are complete and points are evenly distributed. Furthermore, it has the point-to-point correspondences with each other, which means it can be used as our ground truth for quantitative analysis. Also, the SPRING dataset is divided into male and female subsets. To train a model whose muscle and tissue properties are specific to female and male, we separately train male and female shape models. For each gender, 200 meshes from SPRING dataset are used as the training set and the remaining meshes are regarded as the testing set. To show the superior performance of our method, we compare MABR method with NICP in [2], ANICP in [11], and PCA deformation on SPRING dataset. We compute the 3D shape root mean square error (RMS Error) with the Eq. 19 to measure the accuracy of four methods.

$$\text{RMSError} = \sqrt{\frac{\sum_{i=1}^{n}(p_i - \hat{p_i})^2}{n}}, \quad (19)$$

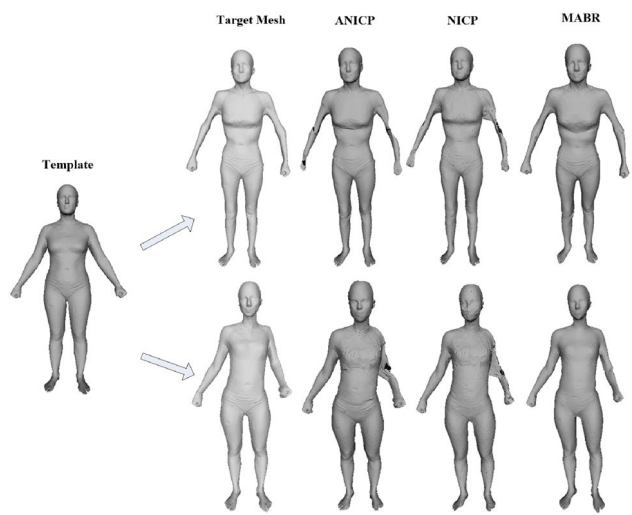

**Fig. 7** The front view of fitted results of ANICP, NICP and MABR in the case that the shape of general template differs a lot from the target mesh

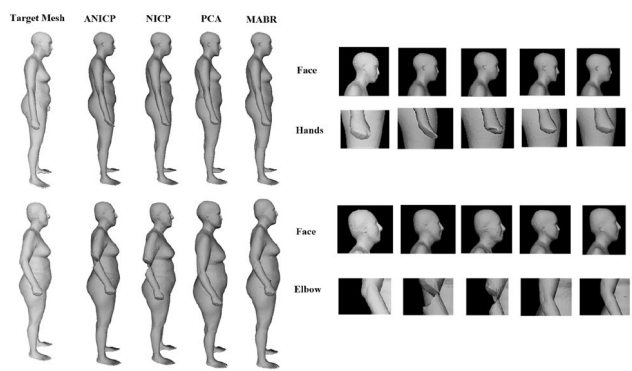

**Fig. 8** The detail comparison of fitting results on SPRING dataset. The side of the raw scan and the fitted results of ANICP (*column 2*), NICP (*column 3*), PCA (*column 4*) and MABR (*column 5*). Besides the comparison of the full body, the details of the face, hand and elbow from each method are also compared subsequently

where points $p_i$ and $\hat{p}_i$ are corresponding points of the ground truth and the fitted results. $n$ is the number of the points in 3D template mesh. As shown in Fig. 6, the accuracy of our method is comparable with ANICP and is much higher than NICP and PCA. In PCA method, the whole model is only controlled by the trained orthogonal basis which cannot cover all the shape variations. Consequently, the accuracy of PCA is the lowest.

Moreover, when the body shape of template is very different from the shape of target, MABR is able to present better fitting results, showing more complete and meaningful limb parts. When the shapes of the template and target scans vary a lot, it will be easier for the icp-based algorithms to find the nearest neighbor incorrectly and more meaningful vertices will be regarded as outliers. We show the front view of some fitted results in Fig. 7, which illustrates the robustness of MABR to outliers. Due to the poor accuracy of PCA method, we do not show the PCA results here. In Fig. 7, we compare the fitting results of two different body shapes from a general template. The first target

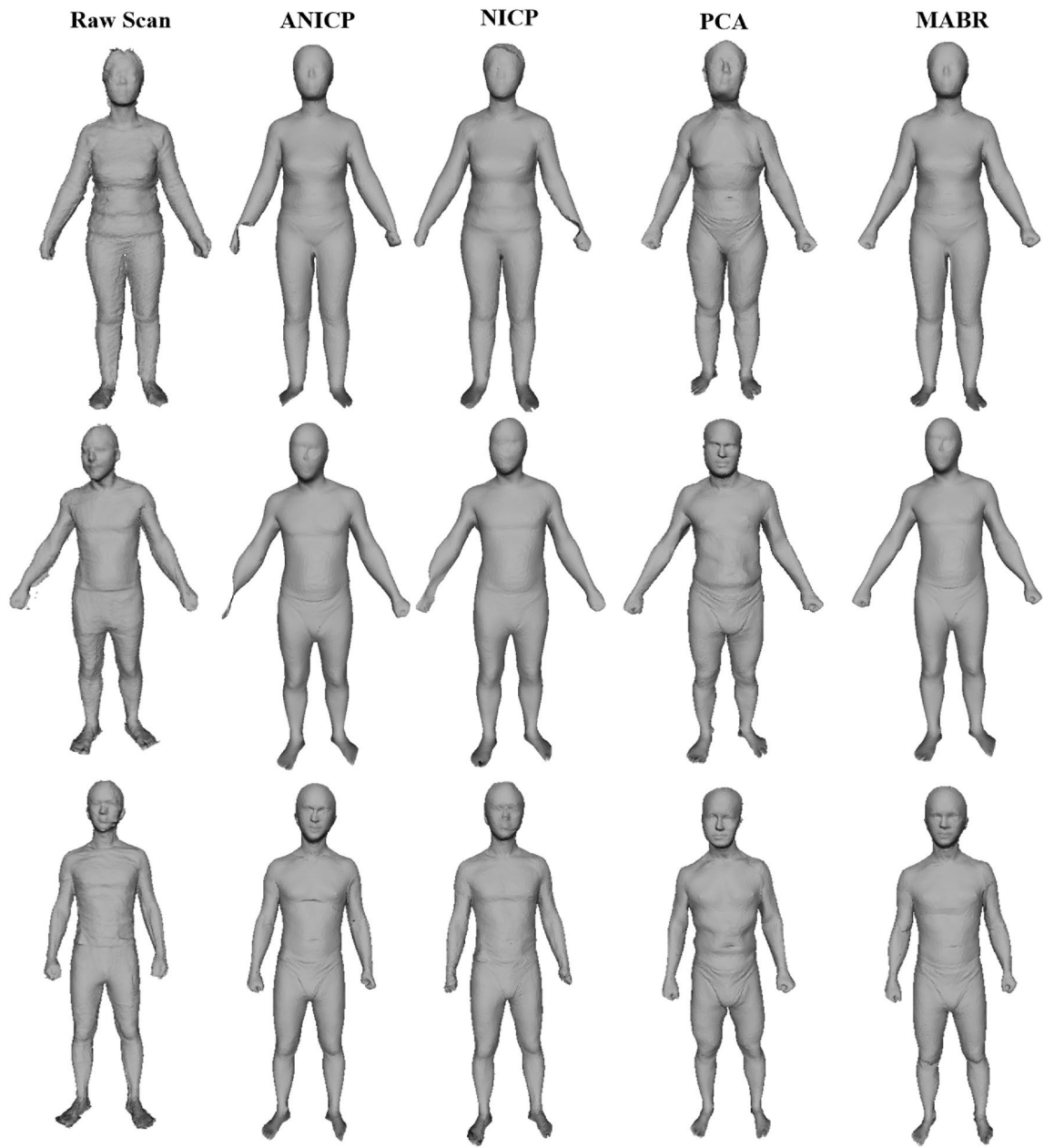

**Fig. 9** Fitting results from Kinect scans. *Column 1* shows the raw body scans, the *second* to the *last columns* illustrate the shapes from ANICP, NICP, PCA and the proposed MABR method, respectively

mesh differs a lot from the template in the arm part and we can see that neither ANICP nor NICP can obtain a complete arm for the given target while the proposed MABR method can not only get a meaningful arm but also make the results similar with the target mesh (which is reflected by the face). What is more, the contour of hands of MABR is clearer than ANICP and NICP. This is shown by the fact that the hole of the fist is visible in MABR results. In the second line, the left arm of the target mesh is bent. Due to the low overlapping degree, the limbs parts are regarded as outliers in ANICP and NICP. As a result, the fitting results of NICP and ANICP in the hand and foot parts are distorted and erroneous while our MABR method successfully fits to the target mesh and keeps complete arm shapes at the same time.

We also compare the details of fitted results from the above four methods in Fig. 8. We can see that for the hand and elbow parts, the PCA method and the proposed MABR method are much better than the other two approaches. In Fig. 8, compared with PCA and MABR, the hand parts of ANICP distorted severely and the fitted hand of NICP is obscure. As for elbow, the results of ANICP and NICP are broken while PCA and MABR are able to preserve the continuity of the fitted mesh. Although PCA can get meaningful results, MABR outperforms it in terms of accuracy, which is reflected by the fitted results in face parts. It can be seen obviously that the face of MABR is much more similar with the raw scan than face of PCA. Basically, MABR successfully recovers the shape of the target mesh. In the elbow part, we can see that the curvature of the mesh from MABR is much closer to the target than PCA's result.

## 5.2 Low-quality scans evaluation

We evaluate the proposed method on low-quality scans which are captured by Microsoft Kinect for XBOX 360. A Kinect is used to scan the person standing at a running turntable from three different heights. The scans are preprocessed to remove background. We compared our proposed MABR method with NICP [2] and ANICP in [11].

Fitting results of these three methods are shown in Fig. 9. It is obvious to see that the proposed MABR method is the only one that models the hand and foot parts completely and, meanwhile, keep high accuracy of the fitting results. We can see that the raw scans have a lot of noises which are close to surface. Large holes exist on top of the head. All these challenges require the registration method to be robust to noises, outliers and holes at the same time. From the results, we can see that neither ANICP nor NICP is not robust enough to obtain complete and accurate registered mesh. The hand parts of ANICP and NICP tend to be distorted and incomplete while the MABR method enables meaningful and complete hands.

The reason for the failure of NICP and ANICP is that, in real scans, the human pose is hard to control so that the limbs are usually not completely overlapped with the template.

Therefore, the shape of the closest points of the limbs cannot keep the limb shape of scans, resulting in unexpected fitted shapes. Since our fitting procedures are active, the limb parts of the template can be stretched along with the direction of PCA basis before performing non-rigid ICP, recovering the size of the hand and foot roughly. In this way, our MABR method is not only able to keep a good shape of the scan but also robust to noises.

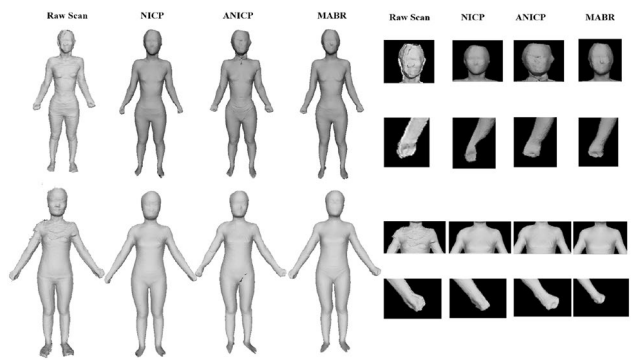

**Fig. 10** The comparison of NICP, ANICP and MABR in the case of hierarchical noises

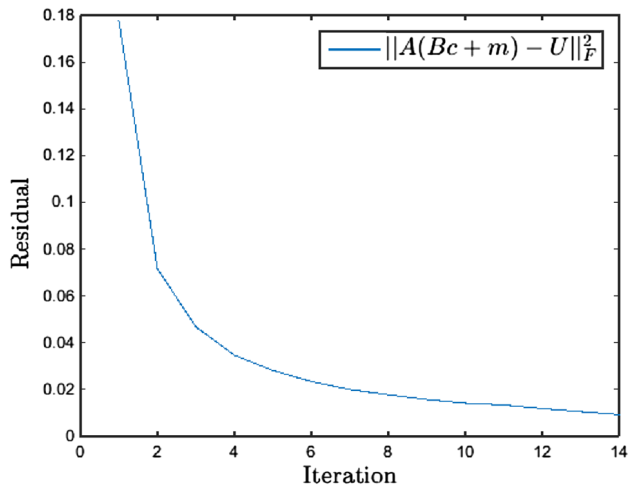

**Fig. 11** Example of residual error changes as the fitting of left hand in the second level progresses

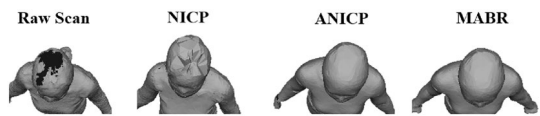

**Fig. 12** The comparison of hole tolerance

**Fig. 13** Examples of K3D-Hub human body scans dataset. We invited both male and female subjects. The ages of subject ranges from 18 to 40. The nationalities of the subjects mainly include Asia and Europe. Each subject performs 5 different poses

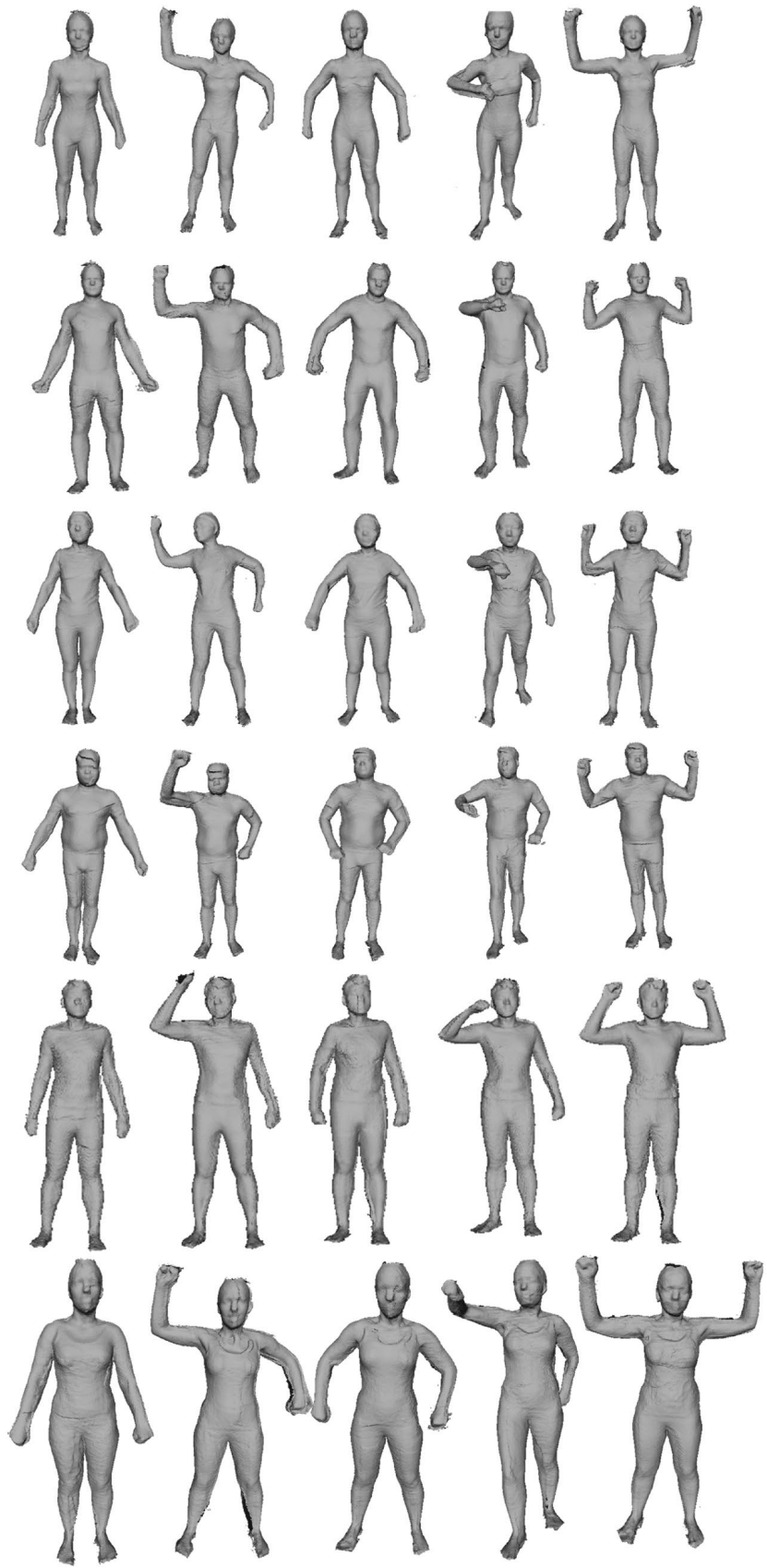

The hierarchical noises are common in the Kinect scans. On one hand, some subtle movements are inevitable when the subjects are trying to keep a certain pose for a few minutes. In this case, hierarchical noises may appear around arms. On the other hand, the subjects are standing on a running turntable. The resulting movements of body may cause hierarchical noises around body surface. In Fig. 10, we compare the fitting results of NICP, ANICP and the proposed MABR in the case of hierarchical noises. We can see that hierarchical noises are distributed on the face, hand, and chest in the raw scans. The fitting results of NICP are ambiguous, without presenting the shape of hands; while, the results of ANICP and MABR can keep hand shapes. The MABR also shows more similar fitting results in faces and hands.

We also show the robustness to holes of MABR in Fig. 12. Even though there exist big holes on top of the head in the raw scan, MABR and ANICP can fill the hole smoothly, which benefits from the training of the prior knowledge. NICP merely relies on finding the nearest points on the target, which is sensitive to holes. Therefore, as illustrated in Fig. 12, the fitted result of NICP is uneven.

In addition, our MABR method has very nice convergence properties. In Fig. 11, we show one example of residual error changes as the fitting of left hand progresses. As can be seen, the residual error monotonically decreases and gradually converges to a minimum value.

All the fitting results in the above experiments are deformed from the same template mesh, the mean shape of the training set. As shown in Figs. 9 and 10, the lifting angles of arms of the test data are not the same and in Fig. 8 the arms of test data are bent, while the arms of template are straight. Arms in these scans tend to be regarded as outliers in NICP and ANICP methods but the proposed method is able to keep the meaningful shape in the registration process, which shows that our method can be applied to scans with different poses in some degree. However, when the target presents different postures with the template like the 2–6 columns shown in Fig. 13, it is still challenging for our method to obtain a reasonable fitting result from the natural standing template for the reason that the searched nearest points cannot keep the shape of hands during fitting. The resulting fitting results will be erroneous. Hence, we reckon that fitting on various poses could be one of our future works.

## 6 Conclusions

In this paper, we propose a multilevel active registration method which combines the non-rigid ICP with the statistical shape model to automatically fit the body template model to the target point clouds. Since the PCA shape model is trained with 200 registered mesh, the combination of PCA makes our method robust to noise, outliers and holes. We have shown that the performance of proposed algorithm is comparable with the state-of-the-art non-rigid registration methods and outperforms them when it comes to the alignment of hands/foot parts. Experiments verify that our approach is robust to both noisy Kinect scans and high-quality meshes. Besides the robust MABR method, a Kinect-based human body dataset, named K3D-Hub, is collected which is the first publicly available low-quality human body scans dataset.

*Limitations* Our registration algorithm manages to register a high-quality template mesh to noisy Kinect scans with similar poses. However, when the initial poses of template and target scans differ much, it is still challenging for our method to generate a reasonable fitting result, particularly in the limb parts. We believe that fitting on various poses will be one of our future work. What is more, our targets are scans captured from subjects with tight clothes and hats. It is still challenging to deform the template to meshes with loose clothes like dresses/skirt. This is because the deformation of loose clothes and hair does not follow the deformation of human body muscle. Unexpected results will appear if we apply our trained morphable model to loose cloths and hair. Moreover, in the future, we plan to speed up the fitting algorithm to support real-time applications.

**Acknowledgements** Funding was provided by China Scholarship Council (Grant No. 201406070079).